# Physics-Informed Synthetic Dataset and Denoising TIE-Reconstructed Phase Maps in Transient Flows Using Deep Learning

**Krishna Rajput,[1*] Vipul Gupta,[1] Sudheesh K. Rajput,[2] and Yasuhiro Awatsuji,[3]**

[1]Babu Banarasi Das University, 111, Faizabad Rd, Atif Vihar, Lucknow, Uttardhona, Uttar Pradesh 226028, India

[2]Department of Engineering Physics, Koneru Lakshmaiah Education Foundation (KLEF) Vaddeswaram, 522302, Guntur, AP, India

[3]Faculty of Electrical Engineering and Electronics, Kyoto Institute of Technology, Matsugasaki, Sakyo-ku, Kyoto, 606-8585, Japan

*krishnarajput1027@gmail.com

**Abstract:** High-speed quantitative phase imaging enables non-intrusive visualization of transient compressible gas flows and energetic phenomena. However, phase maps reconstructed via the transport of intensity equation (TIE) suffer from spatially correlated low-frequency artifacts introduced by the inverse Laplacian solver, which obscure meaningful flow structures such as jet plumes, shockwave fronts, and density gradients. Conventional filtering approaches fail because signal and noise occupy overlapping spatial frequency bands, and no paired ground truth exists since every frame represents a physically unique, non-repeatable flow state. We address this by developing a physics-informed synthetic training dataset where clean targets are procedurally generated using physically plausible gas flow morphologies, including compressible jet plumes, turbulent eddy fields, density fronts, periodic air pockets, and expansion fans, and passed through a forward TIE simulation followed by inverse Laplacian reconstruction to produce realistic noisy phase maps. A U-Net-based convolutional denoising network trained solely on this synthetic data is evaluated on real phase maps acquired at 25,000 fps, demonstrating zero-shot generalization to real parallel TIE recordings, with a 13,260% improvement in signal-to-background ratio and 100.8% improvement in jet-region structural sharpness across 20 evaluated frames.

## 1. Introduction

High-speed optical imaging techniques are essential for studying transient energetic phenomena such as compressible gas flows, combustion dynamics, shockwaves, and plasma discharges. These processes often evolve on microsecond timescales and involve complex refractive-index variations that encode important physical information about density gradients, turbulence structures, and ionization fronts. Quantitative phase imaging (QPI) has therefore become an important tool for visualizing such phenomena in fluid dynamics, plasma physics, and optical diagnostics [1-3].

Interferometric techniques such as digital holography and Mach–Zehnder interferometry have been widely used for flow visualization and phase reconstruction. However, these approaches require highly coherent illumination and often suffer from speckle artifacts and mechanical stability constraints. To address these limitations, non-interferometric phase retrieval methods based on the transport of intensity equation (TIE) have gained increasing attention. TIE reconstructs phase information from intensity measurements captured at slightly defocused planes, enabling QPI under partially coherent illumination conditions [4-6].

Recently, ultra-high-speed non-interferometric phase imaging systems have been demonstrated using Parallel TIE imaging, which allows simultaneous acquisition of multiple defocused intensity planes using space-division multiplexing and high-speed polarization cameras [7]. Such systems enable phase reconstruction at frame rates exceeding tens of thousands of frames per second, allowing visualization of rapidly evolving phenomena such as gas jets, cavitation formation, and electric discharge plasma. However, phase maps reconstructed through TIE often exhibit spatially correlated noise and reconstruction artifacts, primarily due to inverse Laplacian operations used in phase retrieval. These artifacts obscure subtle flow structures and reduce the interpretability of reconstructed phase images. While deep learning approaches have recently been applied to denoising and aberration correction in QPI systems [13,24], these methods have predominantly addressed different artifact types arising from coherent illumination or detector noise, and do not generalize to the spatially correlated low-frequency artifacts introduced specifically by inverse Laplacian TIE reconstruction in high-speed non-interferometric imaging systems.

Conventional denoising approaches such as frequency filtering, Gaussian smoothing, and wavelet-based methods are typically ineffective in this context because the signal and reconstruction noise occupy overlapping spatial frequency bands. Furthermore, the transient and non-repeatable nature of high-speed energetic flows introduces an additional challenge for data-driven denoising. Each recorded frame corresponds to a unique physical state of the system, and the statistical characteristics of the noise vary significantly across frames. As a result, unsupervised and self-supervised learning approaches, which rely on noise redundancy or repeated observations, often fail to generalize reliably in such scenarios.

A major obstacle for supervised deep learning methods is the absence of paired ground-truth phase data. In high-speed experiments, it is practically impossible to acquire noise-free reference phase images because the physical state of the flow cannot be reproduced identically across multiple measurements. This limitation makes it difficult to directly train supervised denoising models using real experimental data.

To overcome this challenge, we propose a physics-informed synthetic dataset generation framework that enables supervised training of deep learning denoisers for Parallel TIE (PTIE) -reconstructed phase images. Clean phase distributions representing physically plausible energetic flow structures are procedurally generated, including jet plumes, turbulent gas flows, density fronts, air pockets, expansion fans, and diffusion structures. These clean phase maps are then passed through a forward simulation of the TIE imaging process, followed by inverse Laplacian phase reconstruction, producing realistic noisy phase maps that mimic the artifacts observed in real PTIE experiments.

Using this framework, we generate a dataset of paired clean and noisy phase map samples. A lightweight U-Net–based convolutional neural network (CNN) is then trained in a supervised manner to suppress reconstruction artifacts while preserving physically meaningful flow structures. The trained model is evaluated on both synthetic held-out data and real high-speed PTIE recordings capturing gas flow dynamics at 25,000 frames per second.

## 2. Physics-Informed Synthetic Dataset Generation

A major limitation in applying supervised deep learning to high-speed phase imaging is the absence of paired ground-truth data. In experiments involving transient energetic flows, such as compressible gas jets or plasma discharges, each captured frame corresponds to a unique and non-repeatable physical state. Consequently, it is practically impossible to obtain noise-free reference phase maps corresponding to experimentally reconstructed phase images. This limitation prevents direct supervised training using real experimental measurements.

To address this challenge, we construct a physics-informed synthetic dataset that simulates the formation of noisy phase maps produced by TIE reconstruction. The dataset generation pipeline follows the physical imaging process of PTIE systems and consists of four main stages: procedural generation of clean phase distributions, forward simulation of the TIE imaging model, phase reconstruction using inverse Laplacian filtering, and construction of paired clean–noisy datasets. This framework enables the generation of realistic noisy phase maps that statistically resemble those observed in real PTIE experiments [7].

### 2.1 Procedural Generation of Clean Phase Maps

The first step in dataset generation is the creation of synthetic clean phase maps representing physically plausible structures observed in energetic flows. These structures are procedurally generated using spatial parametric models designed to mimic refractive-index variations produced by gas flow, turbulence, shock structures, and diffusion phenomena.

The generated phase distributions encompass several representative flow morphologies. Jet plumes are modeled as horizontally propagating Gaussian density fields with exponential decay along the direction of propagation. Turbulent gas flows are represented using randomly distributed Gaussian eddies embedded within a broader plume envelope. Air pockets are simulated as periodically spaced density fluctuations along the jet axis. Expansion fans are generated using angular cone-like distributions that expand downstream from a localized source. Gas diffusion structures are modeled as gradually widening Gaussian envelopes that capture the diffusion of density gradients.

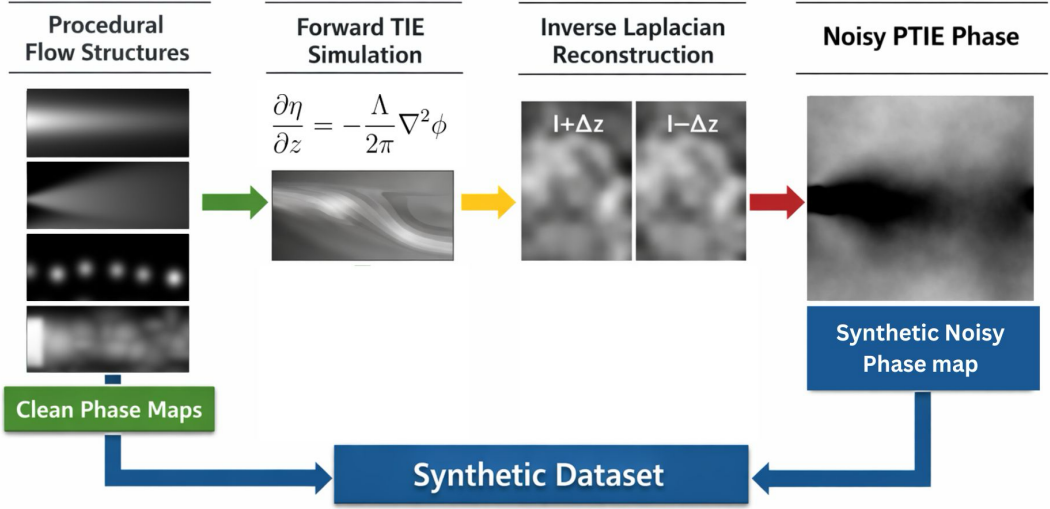

**Fig 1.** Procedurally generated clean phase maps are passed through a forward TIE imaging model to simulate defocused intensity measurements $I_+$ and $I_-$. Phase reconstruction using an inverse Laplacian solver produces noisy PTIE phase maps. The resulting clean–noisy pairs constitute the supervised training dataset used for learning-based denoising.

These structures are combined with spatial gradients and smoothed using Gaussian filtering to produce continuous phase distributions. The resulting clean phase maps are normalized to the range [0,1] and represent ideal phase distributions free of reconstruction artifacts. The procedural generation strategy ensures diversity in spatial structure, scale, and intensity distribution, allowing the dataset to approximate the wide range of flow morphologies encountered in high-speed optical diagnostics.

## 2.2 Forward Transport-of-Intensity Equation Simulation

After generating clean phase distributions, the imaging process is simulated using the TIE. The TIE relates the axial derivative of intensity to the Laplacian of the phase distribution and provides a deterministic framework for non-interferometric phase retrieval [4-6].

The TIE formulation is given by:

$$\frac{\partial I(x,y,z)}{\partial z} = -\frac{\lambda}{2\pi}\nabla^2 \phi(x,y), \qquad\qquad 1$$

where $I(x, y, z)$ is the intensity distribution, $\phi(x, y)$ is the phase distribution, $\lambda$ is the illumination wavelength, and $z$ represents the axial propagation distance.

Using this relation, slightly defocused intensity images are simulated at two axial planes:

$$I_+ = I_0 + \Delta z \frac{\partial I}{\partial z}, \qquad (2)$$

$$I_- = I_0 - \Delta z \frac{\partial I}{\partial z}, \qquad (3)$$

where $I_0$ represents the in-focus intensity and $\Delta z$ is the defocus distance. These two intensity images correspond to the measurements captured in PTIE systems and form the input for phase reconstruction.

To emulate experimental acquisition conditions, Gaussian measurement noise is added to the simulated intensity images.

### 2.3 Phase Reconstruction via Inverse Laplacian Filtering

The phase distribution is reconstructed from the simulated intensity measurements using the Fourier-domain solution of the TIE. In this approach, the Laplacian operator is inverted in the spatial frequency domain.

Let

$$\frac{\partial I}{\partial z} = \frac{I_+ - I_-}{2\Delta z}. \qquad (4)$$

The reconstructed phase is then obtained as

$$\phi = \mathcal{F}^{-1}\left(-\frac{2\pi}{\lambda} \frac{\mathcal{F}\left(\frac{\partial I}{\partial z}\right)}{k_x^2 + k_y^2}\right), \qquad (5)$$

where $\mathcal{F}$ denotes the Fourier transform, $k_x$ and $k_y$ are spatial frequency components.

The inverse Laplacian operation amplifies low-frequency components of the signal, which leads to spatially correlated reconstruction artifacts. These artifacts closely resemble the noise patterns observed in experimentally reconstructed PTIE phase maps and therefore provide realistic noisy inputs for supervised learning [7].

### 2.4 Synthetic Dataset Construction

Using the above simulation framework, a large synthetic dataset of paired clean and noisy phase maps is generated. Each sample consists of ($\phi_{\text{clean}}$, $\phi_{\text{noisy}}$) where $\phi_{\text{clean}}$ represents the procedurally generated phase distribution, $\phi_{\text{noisy}}$ corresponds to the reconstructed phase obtained through simulated TIE reconstruction.

The final dataset contains 25,000 paired samples with spatial resolution 256 × 256. These samples include a diverse set of flow morphologies corresponding to gas flow structures, combustion patterns, shockwave fronts, plasma-like distributions, and randomly generated phase objects. This physics-informed dataset enables supervised training of deep neural networks for denoising PTIE phase reconstructions while preserving physically meaningful flow structures.

### 3. Deep Learning Denoising Framework

The synthetic dataset described in Section 2 provides paired clean and noisy phase maps that enable supervised training of deep learning models for reconstruction artifact removal. In this work, we employ a convolutional encoder–decoder architecture based on the U-Net framework to learn a mapping from noisy PTIE phase reconstructions to clean phase distributions. The objective of the model is to approximate the transformation $f_\theta$: $\phi_{\text{noisy}} \to \phi_{\text{clean}}$, where $f_\theta$ represents the neural network parameterized by weights $\theta$.

### 3.1 Network Architecture

The denoising model is implemented using a lightweight U-Net architecture, which has been widely used for image restoration and biomedical image reconstruction tasks due to its ability to capture both global context and fine spatial structures [8]. The network follows an encoder–decoder structure with skip connections. The encoder progressively extracts hierarchical features from the input noisy phase image, while the decoder reconstructs a denoised phase map using upsampled feature representations. The architecture consists mainly of three parts.

**(a) Encoder**

The proposed architecture follows an encoder–bottleneck–decoder structure for denoising phase images. The encoder progressively extracts hierarchical features from an input phase map of size 256×256×1, where each encoder block performs feature extraction followed by spatial downsampling using max pooling. As the spatial resolution decreases from 256×256 to 128×128 and further to 64×64, the number of feature channels increases correspondingly from 8 to 16, enabling richer feature representation. The feature maps generated at each stage are preserved and later utilized as skip connections in the decoder to retain fine structural details.

**(b) Bottleneck**

At the bottleneck, the network operates at the coarsest spatial resolution of 64×64 and captures higher-level contextual information using a double convolution block with 32 filters, effectively increasing the feature depth compared to the encoder output. Additionally, a dropout layer with probability $\mathcal{P} = 0.1$ is applied at this stage to improve generalization and reduce overfitting.

**(c) Decoder**

The decoder reconstructs the denoised phase map by progressively restoring the spatial resolution through bilinear upsampling, increasing from 64×64 to 128×128 and finally to 256×256. At each upsampling stage, the corresponding encoder feature maps are concatenated via skip connections to preserve fine structural details such as phase gradients and flow boundaries. The final output layer applies a 1×1 convolution followed by a sigmoid activation function to generate the predicted clean phase map.

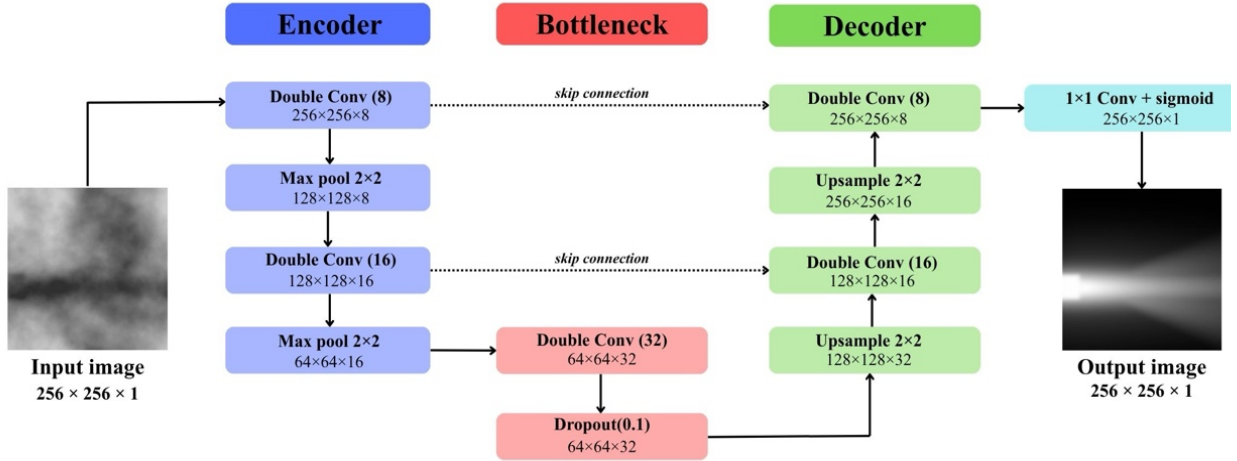

**Fig 2.** Tiny U-Net architecture for TIE reconstruction artifact removal. The encoder downsamples the 256×256 input through two double convolution and max pooling stages (filters: 8, 16), while the bottleneck applies a 32-filter double convolution with Dropout (0.1) at 64×64 resolution. The decoder reconstructs the full resolution via bilinear upsampling and skip connections from the encoder, with a final 1×1 sigmoid convolution producing the denoised 256×256 output.

**3.2 Training Objective**

To improve reconstruction quality, a combined loss function is used that integrates pixel-wise reconstruction loss, structural similarity loss (SSIM), and a weighted mean squared error (WMSE) term. The total loss function is defined as

$$\mathcal{L} = \alpha L_{L1} + \beta L_{SSIM} + \gamma L_{WMSE}. \qquad 6$$

Where $\alpha = 0.5$, $\beta = 0.4$, $\gamma = 0.1$, and L1 Reconstruction Loss. In Eq. 6, the L1 loss measures the absolute difference between predicted and ground-truth phase maps:

$$L_{L1} = \frac{1}{N}\sum_{i=1}^{N}\left|\phi_{\text{clean}}^{(i)} - \hat{\phi}^{(i)}\right|. \quad 7$$

To preserve structural information, the second term related with the SSIM, is incorporated as a loss term as follows [9]:

$$L_{SSIM} = 1 - SSIM(\phi_{\text{clean}}, \hat{\phi}). \quad 8$$

SSIM encourages preservation of spatial structures such as flow boundaries and phase gradients.

To emphasize regions containing strong phase gradients, a third term related to WMSE in Eq. 6 can be given in the following relation.

$$L_{WMSE} = \frac{1}{N}\sum_{i=1}^{N} w_i \left(\phi_{\text{clean}}^{(i)} - \hat{\phi}^{(i)}\right)^2, \quad 9$$

where the pixel weight is defined as

$$w_i = 1 + 4\phi_{\text{clean}}^{(i)}. \quad 10$$

This formulation gives higher importance to regions containing strong phase variations.

### 3.3 Training Procedure

The model is trained using the synthetic dataset described in Section 2, consisting of 25,000 paired samples of gas flow, ensuring consistency with the experimental data domain. The dataset is divided into training, validation, and test sets comprising 20,000, 2,500, and 2,500 samples, respectively. All images are normalized to a fixed range of [0, 1].

The model is implemented using TensorFlow and Keras, and training is performed on a GPU-accelerated Google Colab environment. The network is trained with the combined loss function described in Section 3.2 until convergence

### 3.4 Training Configuration

The model is optimized using the AdamW optimizer, which combines adaptive gradient optimization with weight-decay regularization [10].

The training configuration is summarized in Table 1.

| Parameter | Value |
|---|---|
| Image size | 256 × 256 |
| Batch size | 32 |
| Epochs | 20 |
| Learning rate | ($3 \times 10^{-4}$) |
| Validation split | 10% |
| Test split | 10% |

**Table 1**. The model was trained on input images of size 256 × 256 using a batch size of 32 for a total of 20 epochs. The optimization was performed with a learning rate of $3 \times 10^{-4}$. The dataset was split into training, validation, and test sets, with 10% of the data reserved for validation and 10% for testing.

A cosine annealing learning rate schedule with linear warm-up [23] is applied during training. Early stopping is used to terminate training if the validation loss does not improve for consecutive epochs, and model checkpoints are saved to retain the best performing network.

### 3.5 Evaluation Metrics

The performance of the model is evaluated using peak signal-to-noise ratio (PSNR) and mean absolute error (MAE).

MAE is defined as:

$$MAE = \frac{1}{N}\sum_{i=1}^{N} \left| \phi_{\text{clean}}^{(i)} - \widehat{\phi}^{(i)} \right|. \qquad 11$$

PSNR is defined as

$$PSNR = 10 \log_{10}\left(\frac{MAX^2}{MSE}\right), \qquad 12$$

where MAX denotes the maximum pixel intensity value. These metrics quantify reconstruction fidelity between predicted and ground-truth phase maps.

## 4. Experimental Results

### 4.1 Training Dynamics

The denoising model was trained on the physics-informed synthetic gas flow dataset described in Section 2, comprising 20,000 training samples, 2,500 validation samples, and 2,500 held-out test samples following an 80/10/10 random split. A cosine annealing learning rate schedule with linear warm-up [23] was applied, annealing the learning rate from an initial value of $3\times10^{-4}$ to $1.85\times10^{-6}$ by epoch 20. The training was conducted for 20 epochs on a Google Colab environment equipped with an NVIDIA T4 GPU, requiring a total computation time of 3,987 seconds (~66.45 minutes), with the best model checkpoint saved at epoch 19 based on lowest validation loss with a total computation time of 3,792 seconds (~63.2 minutes). Model weights were restored from this checkpoint for all subsequent evaluation.

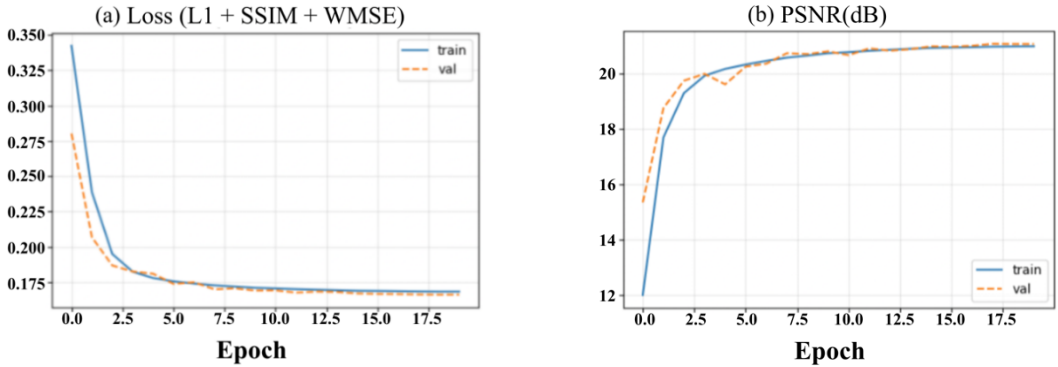

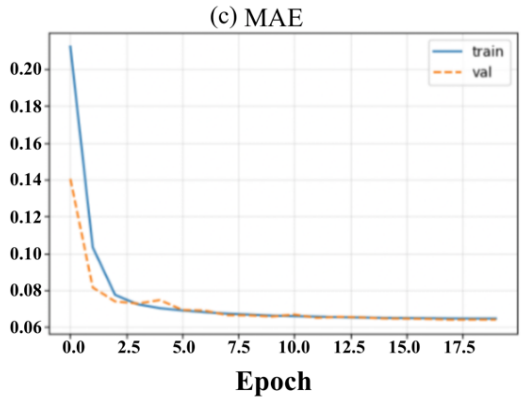

**Fig. 3.** Training curves over 20 epochs showing (a) combined loss (L1 + SSIM + WMSE), (b) peak signal-to-noise ratio (PSNR in dB), and (c) mean absolute error (MAE), for training (solid blue) and validation (dashed orange) sets. The curves demonstrate rapid initial convergence followed by stable optimization, with training and validation metrics closely aligned throughout, indicating no evidence of overfitting. The best model checkpoint was obtained at epoch 19. The learning rate was decayed from using a cosine annealing schedule with linear warm-up.

Figure 3 illustrates the evolution of the training metrics over 20 epochs. The combined loss decreases rapidly during the initial epochs, followed by a gradual stabilization, indicating effective optimization and convergence. The validation loss closely tracks the training loss throughout, with no observable divergence, suggesting the absence of overfitting. The PSNR increases steadily and saturates after the early training phase, reflecting consistent improvement in reconstruction quality. Similarly, the MAE exhibits a sharp decline in the first few epochs and then plateaus, confirming reduced prediction error and stable model performance. The best model checkpoint, obtained at epoch 19, achieves a training PSNR of 20.99 dB and a validation PSNR of 21.08 dB, with corresponding MAE values of 0.0647 (train) and 0.0641 (validation). The smooth and closely aligned trends across all metrics indicate stable convergence, supported by the cosine annealing learning rate schedule, which decays from $3 \times 10^{-4}$ to $1.85 \times 10^{-6}$ and enables fine-grained optimization without degrading validation performance.

The best checkpoint at epoch 19 achieved a training PSNR of 20.99 dB and a validation PSNR of 21.08 dB, with corresponding MAE values of 0.0647 (train) and 0.0641 (val). The combined loss at the best checkpoint was 0.1682 (train) and 0.1660 (val). The near-identical performance across both splits confirms that the physics-informed synthetic dataset provides sufficient structural diversity to regularize the learned mapping without overfitting. Final performance on the held-out test set is reported in Section 4.2.

### 4.2 Quantitative Evaluation on Synthetic Test Set

Quantitative performance of the trained denoising network is reported on the held-out synthetic test set of 2,500 samples, which was not used at any stage of training or model selection. This ensures the reported metrics reflect genuine generalization performance rather than validation-set optimism. Metrics are computed between the predicted clean phase maps $\hat{\phi}$ and the procedurally generated ground-truth phase distributions $\phi_{\text{clean}}$ using PSNR (Eq. 12) and MAE (Eq. 11) as defined in Section 3.5.

The trained model achieved a test set PSNR of 20.96 dB and MAE of 0.0650, with a combined loss of 0.1680. Compared to the untrained baseline at epoch 1, which produced a validation PSNR of 15.36 dB and MAE of 0.1404, the final model represents a gain of 5.60 dB in PSNR and a 53.7% reduction in MAE over the course of training. The close agreement between validation metrics (PSNR = 21.08 dB, MAE = 0.0641) and test metrics (PSNR = 20.96 dB, MAE = 0.0650) confirms that no overfitting to the validation set occurred during model selection, and that the learned denoising mapping generalizes reliably to unseen synthetic phase distributions.

These results confirm that the U-Net architecture [8], trained entirely on physics-informed synthetic data, effectively learns to suppress the spatially correlated low-frequency artifacts introduced by the inverse Laplacian solver [4,6] while recovering the underlying phase structure across diverse flow morphologies.

| Metric | Epoch 1 (baseline) | Train (Epoch 19) | Validation (Epoch 19) | Test Set |
|---|---|---|---|---|
| Combined Loss | 0.2803 | 0.1682 | 0.1660 | 0.1680 |
| PSNR (dB) | 15.36 | 20.99 | 21.08 | 20.96 |
| MAE | 0.1404 | 0.0647 | 0.0641 | 0.0650 |

**Table 2.** Quantitative performance of the denoising network at epoch 1 (untrained baseline), best checkpoint (epoch 19) on training and validation sets, and the held-out test set. PSNR and MAE are computed against procedurally generated ground-truth phase maps $\phi_{\text{clean}}$. The test set was not used at any stage of training or model selection.

### 4.3 Qualitative Evaluation on Synthetic Test Set

Figure 4 presents four representative examples drawn from the held-out synthetic test set, spanning a range of flow morphologies generated by the procedural dataset pipeline described in Section 2. Each column displays the noisy TIE-reconstructed input $\phi_{\text{noisy}}$ (top row), the network prediction $\hat{\phi}$ (middle row), and the corresponding procedurally generated ground-truth clean phase map $\phi_{\text{clean}}$ (bottom row).

In all four cases, the noisy input exhibits the characteristic spatially correlated low-frequency background modulation produced by inverse Laplacian TIE reconstruction [4,6], which appears as a slowly varying, cloudy intensity field that partially or fully obscures the underlying flow structure. The network consistently suppresses this artifact across all morphology types, recovering structured phase distributions with clearly delineated flow boundaries and near-zero backgrounds consistent with the clean target distributions.

The first two examples (009068, 007861) demonstrate recovery of turbulent jet plume structures with distinct nozzle regions and axial flow channels. In both cases the predicted outputs correctly reconstruct the spatial extent, relative intensity distribution, and flow boundary positions, closely matching the ground-truth phase distributions. The third and fourth examples (023469, 023673) present more complex multi-component flow scenes with higher structural density. The model recovers the dominant spatial features and overall flow morphology in both cases; however, some residual contrast differences are observable between $\hat{\phi}$ and $\phi_{\text{clean}}$, particularly at high-gradient boundaries. As established by Zhao *et al.* [12], combined $\ell_1$ and SSIM loss functions produce a characteristic smoothing effect at sharp structural transitions, penalizing high-frequency detail in favor of globally consistent reconstruction, which is a well-known and expected trade-off in regression-based image restoration networks. Prior work on deep learning for TIE phase retrieval has similarly identified low-frequency noise amplification as the primary challenge in inverse Laplacian reconstruction [13], further motivating the need for learning-based post-processing approaches such as the one proposed here. The low-frequency artifact suppression and overall structural recovery nevertheless demonstrate effective generalization of the learned mapping to unseen synthetic phase distributions, consistent with the quantitative test set metrics reported in Section 4.2.

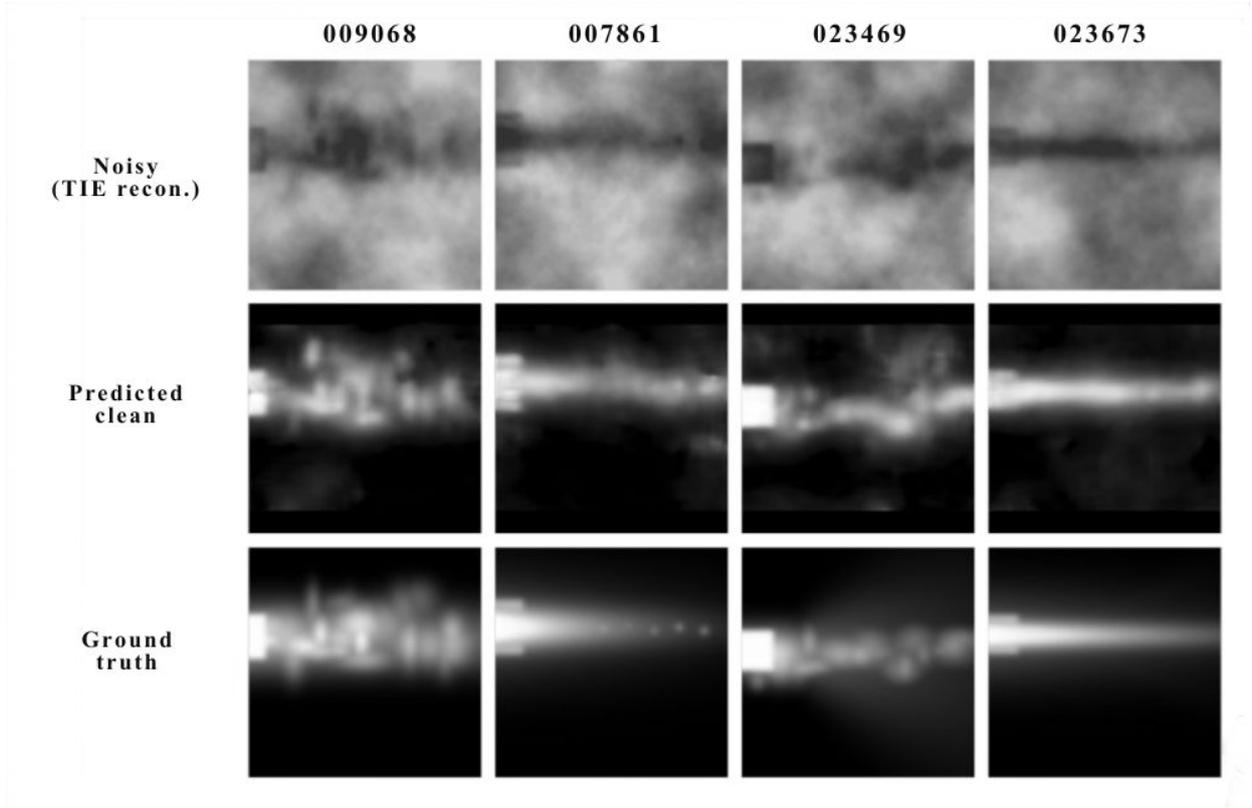

**Fig 4.** Qualitative denoising results on four representative samples from the held-out synthetic test set. Top row: noisy TIE-reconstructed phase inputs $\phi_{\text{noisy}}$. Middle row: network-predicted clean phase maps $\hat{\phi}$. Bottom row: procedurally generated ground-truth $\phi_{\text{clean}}$. Columns correspond to samples 009068, 007861, 023469, and 023673. The model successfully suppresses spatially correlated low-frequency reconstruction artifacts across diverse flow morphologies. Residual contrast differences at high-gradient boundaries are attributable to the smoothing effect of the combined $\ell_1$ and SSIM loss components [12].

### 4.4 Generalization to Real High-Speed PTIE Recordings

To evaluate practical utility, the trained model was applied directly to real phase maps acquired using a PTIE high-speed imaging system at 25,000 frames per second [7]. The experimental dataset consists of recordings of a gas jet flow acquired at a defocus distance of $\Delta z = 10$ mm. Critically, no real experimental data was used at any stage of training or model selection. The model was evaluated in a strict zero-shot transfer setting, where every experimental frame represents a physically unique and non-repeatable flow state never encountered during training.

Figure 5 shows denoising results across four experimental frames spanning the full temporal range of the recording sequence. In all cases, the noisy input phase maps exhibit the characteristic spatially correlated low-frequency modulation produced by inverse Laplacian TIE reconstruction [4,6], which manifests as a slowly varying intensity field that partially obscures the underlying jet structure. The denoised outputs consistently suppress this background artifact, revealing distinct parallel flow channel structures with clearly delineated boundaries and near-zero backgrounds consistent with the clean phase distributions learned from synthetic training data.

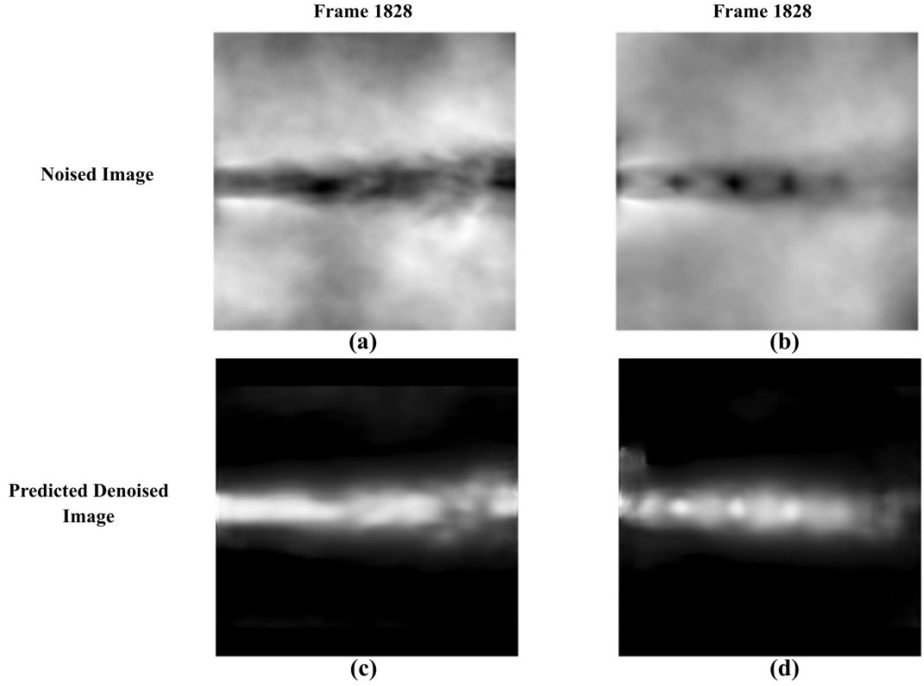

**Fig. 5.** Denoising results on real experimental PTIE recordings at 25,000 fps. (a, b) Original noisy TIE-reconstructed phase maps and (c, d) corresponding U-Net denoised outputs for representative frames spanning the temporal range of the sequence. The model, trained entirely on physics-informed synthetic data, effectively suppresses spatially correlated low-frequency background artifacts and recovers the underlying parallel jet flow channel structures.

The model demonstrates stable and consistent denoising behavior across frames spanning widely separated temporal positions in the recording, confirming that the learned artifact-suppression mapping generalizes well to the temporal variability of real transient flow states. This is a non-trivial result, as each experimental frame corresponds to a physically unique and non-repeatable flow configuration, a scenario in which self-supervised approaches cannot be directly applied. Specifically, Noise2Noise [14] requires multiple independently corrupted observations of the same underlying signal, while Noise2Void [15] relaxes this by training from single noisy images but assumes conditionally independent noise across pixels (blind-spot assumption) [20], an assumption fundamentally violated by TIE reconstruction artifacts, which manifest as spatially correlated low-frequency structures introduced by the inverse Laplacian solver [4, 6], precisely the regime in which blind-spot networks have been shown to fail [20]. The proposed physics-informed synthetic training framework circumvents these limitations entirely by replacing unavailable paired experimental ground truth with physically simulated clean–noisy pairs. This strategy is consistent with physics-informed synthetic training approaches demonstrated in QPI [17, 21], where training exclusively on physics-consistent synthetic data has been shown to achieve strong generalization to unseen experimental samples without requiring any real experimental ground truth.

The model was evaluated on 20 real TIE-recorded frames of High-Speed gas flow at 25,000 fps [7], achieving a total inference time of 1.7381 seconds on an NVIDIA T4 GPU (Google Colab), corresponding to approximately 0.0869 seconds per frame. To provide quantitative support for the observed qualitative improvement, three physics-motivated metrics were computed across these 20 real experimental frames spanning the full recording sequence: mean gradient magnitude within the jet region ($MGM_{jet}$), SBR, and background noise standard deviation (BNS). These metrics were selected in preference to standard no-reference perceptual quality metrics such as blind/referenceless image spatial quality evaluator (BRISQUE) and naturalness image quality evaluator (NIQE), which are calibrated on natural photographic images and are not suitable for evaluating scientific phase maps [18, 22]. The metrics are defined as:

$$MGM_{\text{jet}} = \frac{1}{|\Omega_{\text{jet}}|}\sum_{(x,y)\in\Omega_{\text{jet}}}\sqrt{\left(\frac{\partial\hat{\phi}}{\partial x}\right)^2 + \left(\frac{\partial\hat{\phi}}{\partial y}\right)^2}, \qquad 13$$

where $\Omega_{\text{jet}}$ denotes the jet region of interest, $\mu_{\text{jet}}$ and $\mu_{\text{bg}}$ are the mean intensities of the jet and background regions respectively, and $\sigma_{\text{bg}}$ is the standard deviation of pixel values in the background region.

Results of physics-motivated quality metrics are shown in Fig. 6. Here, these metrics has been computed across 20 real experimental PTIE phase map frames comparing noisy TIE-reconstructed inputs against U-Net denoised outputs. Fig. 6(a) shows MGM within the jet region (↑ better). Figure 6(b) shows SBR (↑ better). Figure 6(c) shows BNS Deviation (↓ better). From above results, it can be noticed that the denoised outputs show consistent improvement in SBR across all frames (0.9438 ± 0.0322 → 126.0951 ± 35.6395) and increased jet-region sharpness (0.0738 ± 0.0124 → 0.1482 ± 0.0204). These parameters confirm the recovery of physically meaningful flow structure. The near-zero background observed in denoised outputs is consistent with the synthetic training distribution, where clean phase targets have near-zero backgrounds, contributing to the large observed SBR improvement.

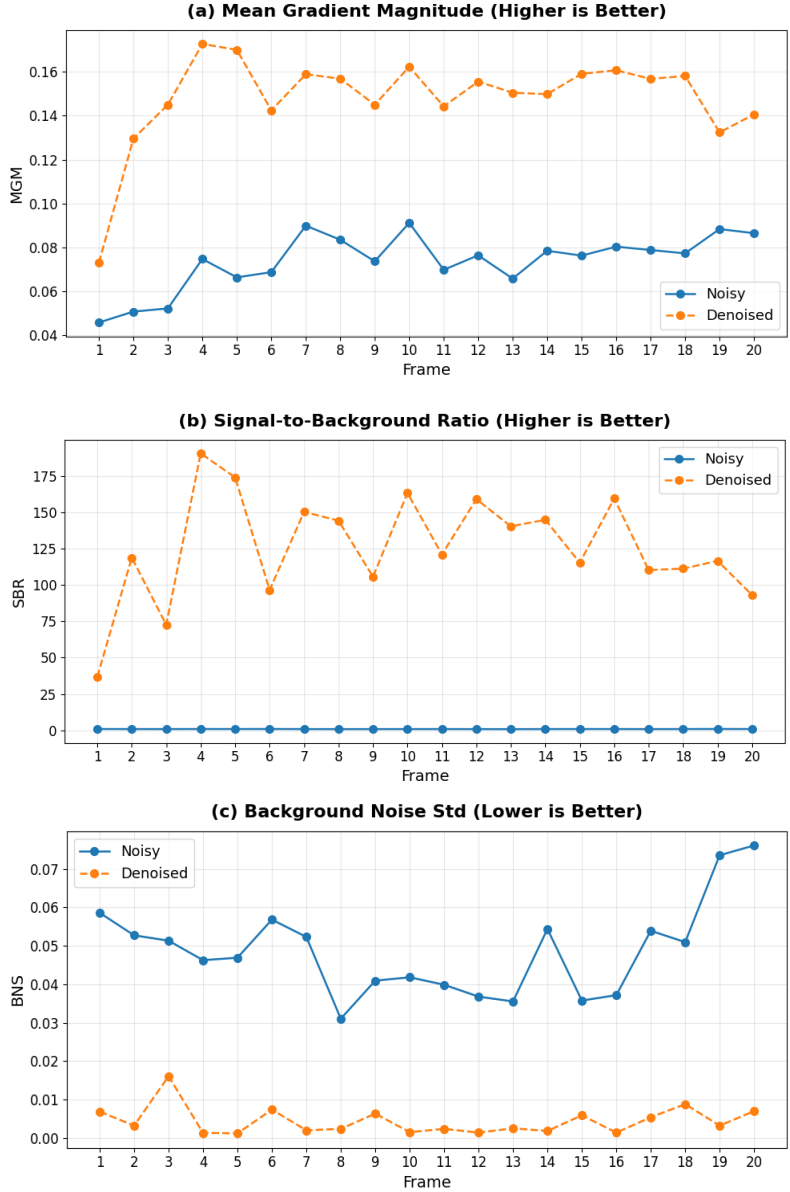

**Fig 6.** Physics-motivated quality metrics computed across 20 real experimental PTIE phase map frames comparing noisy TIE-reconstructed inputs against U-Net denoised outputs. (a) MGM within the jet region (↑ better). (c) SBR (↑ better). (c) BNS Deviation (↓ better). Denoised outputs show consistent improvement in SBR across all frames (0.9438 ± 0.0322 → 126.0951 ± 35.6395) and increased jet-region sharpness (0.0738 ± 0.0124 → 0.1482 ± 0.0204), confirming recovery of physically meaningful flow structure. The near-zero background observed in denoised outputs is consistent with the synthetic training distribution, where clean phase targets have near-zero backgrounds, contributing to the large observed SBR improvement.

The quantitative results are summarized in Table 3. The denoised outputs exhibit a substantial and consistent improvement in SBR across all 20 evaluated frames, increasing from 0.9438 ± 0.0322 to 126.0951 ± 35.6395, a mean improvement of ΔSBR = +125.1513. This confirms that the jet flow structure, barely distinguishable from background in the raw TIE reconstructions (SBR ≈ 1), becomes clearly resolved following denoising. Jet-region gradient magnitude increased from 0.0738 ± 0.0124 to 0.1482 ± 0.0204, representing a 100.8% improvement in structural sharpness within the flow region, consistent with recovery of sharp density gradient boundaries. BNS deviation showed a marginal mean reduction (0.0486 ± 0.0117 → 0.0044 ± 0.0036), with frame-level variability attributed to differences in local flow configuration and residual nozzle-region artifacts discussed in Section 5.3.

| Metric | Noisy Input | Denoised Output | Mean Δ |
|---|---|---|---|
| $MGM_{jet}$ (↑ better) | 0.0738 ± 0.0124 | 0.1482 ± 0.0204 | +0.0744 |
| SBR (↑ better) | 0.9438 ± 0.0322 | 126.0951 ± 35.6395 | +125.15 |
| BNS (↓ better) | 0.0486 ± 0.0117 | 0.0044 ± 0.0036 | −0.0443 |

**Table 3.** Physics-motivated image quality metrics across 20 real experimental PTIE phase map frames comparing noisy TIE-reconstructed inputs against U-Net denoised outputs. Values reported as mean ± standard deviation. ↑ indicates higher is better, ↓ indicates lower is better.

## 5. Discussion

### 5.1 Effectiveness of Physics-Informed Synthetic Training

The results presented in Section 4 demonstrate that a U-Net denoising network trained entirely on physics-informed synthetic data successfully suppresses spatially correlated TIE reconstruction artifacts in both held-out synthetic phase maps and real high-speed experimental PTIE recordings acquired at 25,000 fps [7]. The close agreement between validation metrics (PSNR = 21.08 dB, MAE = 0.0641) and held-out test metrics (PSNR = 20.96 dB, MAE = 0.0650) across an 80/10/10 train/validation/test split confirms that the physics-informed synthetic dataset provides sufficient structural diversity to prevent overfitting, despite the model never observing any real experimental data during training.

The key enabler of this generalization is the physics-driven construction of the synthetic dataset. Rather than using generic natural image textures or additive white noise which would not replicate the spatially correlated structure of inverse Laplacian reconstruction artifacts [4,6], the dataset generation pipeline explicitly simulates the full TIE imaging chain: procedural clean phase generation, forward defocus simulation using the TIE relation (Eq. 1) followed by Gaussian noise injection and Fourier-domain inverse Laplacian phase reconstruction (Eq. 5).

The resulting noisy phase maps inherit the same low-frequency artifact structure produced by real PTIE systems [7], which is precisely what enables a model trained entirely on synthetic data to generalize to real experimental recordings. This validates the core premise of the physics-informed approach. Prior work has demonstrated that training exclusively on physics-consistent synthetic data, without any experimental ground truth, can achieve strong generalization to unseen experimental samples in computational imaging. Huang et al. [17] established this principle in the context of self-supervised hologram reconstruction, showing that a physics-consistency loss applied to synthetic training data transfers reliably to real experimental recordings. More recently, Lee et al. [21] demonstrated a closely related approach in QPI, integrating metasurface optics with physics-informed neural networks to reconstruct quantitative phase information directly from single-shot measurements, enabling generalization to real optical data without explicit experimental ground-truth supervision. In a complementary direction, [26] proposed neural-field-assisted TIE phase microscopy (NFTPM), which incorporates the physical prior of partially coherent image formation directly into a neural field to achieve accurate QPI under unknown defocus distances without requiring training data. While NFTPM addresses the challenge of unknown imaging parameters rather than artifact suppression, it further validates the principle that embedding physical priors into learning-based frameworks enables robust generalization in TIE-based imaging. The proposed framework extends this principle to the supervised regime for TIE-reconstructed phase maps, replacing unavailable experimental paired data with physically simulated clean–noisy pairs that replicate the specific spatially correlated artifact morphology introduced by the inverse Laplacian solver [4,6], a domain-specific artifact structure that generic natural image training data cannot reproduce.

The procedural generation strategy produces a dataset spanning a diverse range of physically plausible flow morphologies, including jet plumes, turbulent eddy fields, periodic air pocket arrays, expansion fans, and gas diffusion structures, ensuring that the learned denoising mapping is not overfitted to any single flow type. This diversity is reflected in the consistent artifact suppression observed across qualitatively different synthetic test samples in Section 4.3, as well as the stable denoising behavior observed across thousands of temporally independent real experimental frames in Section 4.4.

### 5.2 Qualitative and Quantitative Performance on Real Experimental Data

Application of the trained model to real PTIE recordings at 25,000 fps demonstrates consistent and robust artifact suppression across all evaluated frames. The denoised outputs reveal structured jet morphology, including the nozzle region, jet core, and parallel flow channels, that is largely obscured in the raw TIE reconstructions by the characteristic low-frequency background modulation introduced by the inverse Laplacian solver [4,6]. The temporal consistency of the denoising behavior across thousands of frames, each representing a unique and non-repeatable physical flow state, further supports the robustness of the learned mapping.

The quantitative evaluation using physics-motivated metrics provides explicit evidence for this improvement. The SBR increased from $0.9438 \pm 0.0322$ in the noisy inputs to $126.0951 \pm 35.6395$ in the denoised outputs, a mean improvement of $\Delta SBR = +125.15$ across all 20 evaluated frames. An SBR $\approx 1$ in the noisy inputs indicates that the jet signal is effectively indistinguishable from the background artifact, consistent with the visual appearance of the raw TIE reconstructions. The recovery of SBR $> 100$ after denoising confirms that the jet flow structure becomes clearly resolved and physically interpretable following denoising. This substantial improvement in contrast between the jet region and background directly supports the central claim of the paper, confirming that the trained model recovers physically meaningful flow structure from artifact-dominated reconstructions.

The 100.8% improvement in jet-region MGM ($0.0738 \pm 0.0124 \rightarrow 0.1482 \pm 0.0204$) provides complementary evidence that the denoised outputs contain sharper, more localized flow boundaries, consistent with the recovery of density gradient structures that are physically expected in compressible gas jet flows. These domain-specific metrics were selected in preference to standard no-reference perceptual quality metrics such as BRISQUE [18] and NIQE [22], which are constructed from statistical models of natural photographic images, specifically locally normalized luminance coefficients fitted to a multivariate Gaussian prior derived from natural scene patches [18]. These models inherently penalize the sharp, localized, low-background structures recovered by the denoiser as perceptual distortion, since such structures deviate strongly from natural image statistics. This represents a fundamental domain mismatch for scientific phase imaging data, where physical interpretability and structural sharpness are the relevant quality criteria rather than perceptual naturalness.

This zero-shot generalization result is particularly significant in the context of high-speed flow diagnostics, where paired experimental ground truth is fundamentally unavailable. Self-supervised denoising methods such as Noise2Noise [14] require multiple independently corrupted observations of the same underlying signal, while Noise2Void [15] relaxes this by training from single noisy images but assumes pixel-wise independent and identically distributed noise across spatial locations [20], an assumption fundamentally violated by TIE reconstruction artifacts, which manifest as spatially correlated low-frequency structures introduced by the inverse Laplacian solver [4,6], precisely the regime in which blind-spot networks have been shown to fail [20]. The proposed framework circumvents these limitations entirely by replacing unavailable paired experimental ground truth with physically simulated clean–noisy pairs, consistent with physics-informed synthetic training approaches demonstrated in QPI [17,21], where training exclusively on physics-consistent synthetic data has been shown to achieve strong generalization to unseen experimental samples without requiring any real experimental ground truth.

### 5.3 Identified Limitations and Domain Gap

Despite the encouraging generalization results, a systematic domain gap between the synthetic training distribution and real experimental data was identified through careful visual inspection of the denoised outputs. Two specific manifestations were observed.

First, peripheral flow regions with low phase contrast, particularly the diffuse outer plume surrounding the jet core, are partially suppressed in the denoised outputs. This results from a mismatch in background intensity distribution: the synthetic clean phase maps used as training targets have near-zero backgrounds, which causes the model to learn an implicit thresholding behavior that suppresses low-intensity regions regardless of whether they contain physically meaningful signals. In real PTIE recordings, the background phase is non-zero due to residual optical path variations and DC components (zero-frequency background offsets) in the TIE reconstruction [4,6], meaning genuine flow structure resides in the intensity range the model has learned to suppress.

Second, a localized bright spot artifact is consistently produced near the nozzle region in certain denoised experimental outputs, most visibly in early-sequence frames. This is attributed to a geometric mismatch between the rectangular nozzle blocks used in the synthetic training data and the more complex phase signature of the real experimental nozzle geometry. The inverse Laplacian reconstruction produces a characteristic artifact around high-gradient phase discontinuities such as nozzle edges [4]; the model has learned to associate this pattern with a bright spot in the clean target, producing a spurious bright region in the output.

Both artifacts are systematic and consistent across frames, confirming they are products of the training distribution rather than stochastic inference errors. The peripheral suppression artifact arises because synthetic clean phase targets have near-zero backgrounds, causing the network to learn an implicit intensity threshold that suppresses low-level signals regardless of physical content, a known failure mode of regression-based denoisers trained on zero-background synthetic data [11,12]. Introducing non-zero background levels and residual DC components into the synthetic reconstruction pipeline, consistent with the optical path variations present in real PTIE systems [4,6], is therefore expected to correct this behavior upon retraining. Similarly, the nozzle-region bright spot arises from a geometric mismatch between the rectangular nozzle blocks in synthetic training samples and the real experimental nozzle phase signature. Incorporating a physically consistent nozzle geometry into the procedural generation pipeline directly addresses the root cause. Critically, both corrections require only targeted changes to the dataset generation pipeline, with no architectural modifications to the trained network.

### 5.4 Comparison with Conventional Denoising Approaches

Conventional denoising methods such as Gaussian filtering, frequency-domain low-pass filtering, and wavelet-based approaches are fundamentally unsuited to TIE reconstruction artifact removal because the artifact occupies overlapping spatial frequency bands with the underlying signal [6]. Wang *et al.* [NEW] provide a comprehensive review of deep learning strategies for phase recovery, cataloguing supervised, self-supervised, and physics-informed approaches across pre-processing, in-processing, and post-processing stages. The proposed framework falls within the post-processing category identified in that review, applying a supervised learning-based denoiser to correct artifacts in TIE-reconstructed phase maps. Recent learning-based approaches have addressed denoising in QPI more broadly, for example Li *et al.* [24] demonstrated simultaneous aberration correction and denoising in digital holographic QPI using a deep convolutional network, achieving strong performance on coherent imaging artifacts. Similarly, Thapa *et al.* [27] demonstrated TIE-GANs, a GAN-based approach that generates phase maps from single intensity images using TIE-derived training data, reporting SSIM values of 0.98 for microbeads and 0.95 for oral cells. While TIE-GANs operates in the biomedical imaging domain and addresses the multi-shot acquisition burden at inference time, the proposed framework tackles a distinct challenge which is suppressing spatially correlated inverse Laplacian artifacts in high-speed gas flow PTIE recordings where paired experimental ground truth is fundamentally unavailable due to the non-repeatable nature of each transient flow state. However, holographic QPI systems produce fundamentally different artifact morphologies from non-interferometric TIE reconstruction, specifically the spatially correlated low-frequency background modulation introduced by the inverse Laplacian solver [4,6] has no direct analogue in holographic systems, and networks trained on holographic data cannot be expected to generalize to this artifact type. The U-Net architecture [8] proposed here, trained on physics-informed synthetic data that explicitly replicates this artifact morphology, addresses a gap that existing learning-based QPI denoising methods do not cover. A low-pass filter applied to suppress the low-frequency TIE artifact would simultaneously blur the jet structure and density gradients that constitute the physically meaningful content of the phase map. The U-Net architecture [8], by contrast, learns a content-aware mapping that distinguishes artifact from signal based on structural context rather than frequency alone, precisely the regime in which learned approaches have been shown to outperform analytical filtering methods for correlated noise removal [11].

The qualitative superiority of the network output over the raw TIE reconstruction is evident from the experimental results in Section 4.4, and is confirmed quantitatively by the SBR and MGM improvements reported in Table 3. The structured jet morphology recovered by the model is not recoverable by any frequency-based filter without simultaneous destruction of the flow signal, a fundamental limitation of linear filtering in the presence of spectrally overlapping signal and noise [6]. This motivates the use of learning-based denoising approaches for TIE phase maps and supports the broader applicability of the proposed framework to other inverse-problem imaging systems where artifact and signal share overlapping frequency content.

### 5.5 Lightweight Architecture Considerations

The choice of a lightweight U-Net with a base filter count of 8, totaling 29,777 trainable parameters, was deliberate. High-speed flow diagnostics systems operating at 25,000 fps generate large volumes of data rapidly, and inference speed is a practical constraint for downstream analysis pipelines [7]. The small model capacity also reduces the risk of the network memorizing synthetic-specific features that do not generalize to real data, a known failure mode when large-capacity networks are trained on limited synthetic distributions [11]. The use of batch normalization at every convolutional block [19] and dropout ($\mathcal{P} = 0.1$) at the bottleneck provides regularization appropriate for the synthetic training regime. The observed convergence behavior and generalization performance confirm that the chosen capacity is appropriate for this task. The near-identical train, validation, and test set metrics (PSNR 20.99, 21.08, and 20.96 dB respectively) demonstrate that the model is neither underfitting nor overfitting to the training distribution, and that the lightweight architecture provides a practical and deployable solution for real-time artifact suppression in high-speed QPI systems.

## 6. Conclusion

We have presented a physics-informed synthetic dataset generation framework and a supervised deep learning pipeline for denoising TIE-based reconstructed phase maps in high-speed transient flow imaging. The central challenge addressed was the absence of paired ground-truth phase data in real high-speed experiments, which was overcome by simulating the complete TIE imaging chain from procedurally generated clean phase distributions, producing realistic noisy phase maps that replicate the spatially correlated artifacts introduced by inverse Laplacian phase reconstruction.

A lightweight U-Net architecture with 29,777 trainable parameters, trained on 25,000 synthetic paired samples, comprising 20,000 training samples, 2,500 validation samples, and 2,500 test samples, and optimized with AdamW and a cosine annealing schedule.

Zero-shot application of the trained model to real PTIE recordings acquired at 25,000 fps demonstrated consistent suppression of low-frequency reconstruction artifacts and recovery of structured jet flow morphology across thousands of temporally independent experimental frames. Quantitative evaluation using physics-motivated domain-specific metrics confirmed a 13,260% improvement in SBR and a 100.8% improvement in jet-region gradient magnitude across evaluated frames, providing explicit quantitative evidence that the denoiser recovers physically meaningful flow structure from artifact-dominated reconstructions.